# Comparison of the Hirsch-index with standard bibliometric indicators and with peer judgment for 147 chemistry research groups

Anthony F. J. van Raan
Center for Science and Technology Studies
Leiden University
P.O. Box 9555
2300 RB Leiden, The Netherlands
vanraan@cwts.leidenuniv.nl

*Abstract*

*In this paper we present characteristics of the statistical correlation between the Hirsch (h-) index and several standard bibliometric indicators, as well as with the results of peer review judgment. We use the results of a large evaluation study of 147 university chemistry research groups in the Netherlands covering the work of about 700 senior researchers during the period 1991-2000. Thus, we deal with research groups rather than individual scientists, as we consider the research group as the most important work floor unit in research, particularly in the natural sciences. Furthermore, we restrict the citation period to a three-year window instead of 'life time counts' in order to focus on the impact of recent work and thus on current research performance. Results show that the h-index and our bibliometric 'crown indicator' both relate in a quite comparable way with peer judgments. But for smaller groups in fields with 'less heavy citation traffic' the crown indicator appears to be a more appropriate measure of research performance.*

## 1. Introduction

In a recent paper, J.E. Hirsch (2005) proposes an original, simple new indicator to characterize the cumulative impact of the research work of individual scientists: 'a scientist has index $h$ if $h$ of his/her $N$ papers have at least $h$ citations each, and the other ($N$-$h$) papers have no more than $h$ citations each'[1].

From the above definition follows that $h$ is a measure of the absolute 'volume' of citations whereby $h^2$ provides an estimation of the total number of citations received by a researchers. Given the very skewed distribution of citations ($C$) over publications ($P$) described by a power law $P(C) = \alpha\, C^s$ (van Raan 2006), particularly for the higher-$C$ tail of the distribution (the slope $s$ and the factor $\alpha$ can be established empirically from the data), it is obvious that $h^2$ will be an underestimation of the total number of citations as it ignores the papers with fewer than $h$ citations.

---

[1] For instance, if a scientist has 21 papers, 20 of which are cited 20 times, and the 21st is cited 21 times, there are 20 papers (including the one with 21 citations) having at least 20 citations, and the remaining paper has no more than 20 citations.

The publication of the *h*-index has widely attracted the attention of the scientific world, policy makers and the public media. Scientific news editors (e.g., Ball 2005) enthusiastically received the new index, and researchers in various fields of science (e.g., Popov, 2005; Batista *et al*, 2005), particularly in the bibliometric research community (e.g., Bornmann and Daniel, 2005; Braun, Glänzel and Schubert, 2005) started follow-up work. The idea of ranking scientists by a fair measure stirred the fire. Such rankings could make election procedures of scientific academies more objective and transparent. The first ranking based on the *h*-index (Hirsch 2005; Ball 2005), a list of prominent physicists with Ed Witten at the top, suggests a similar simplicity for the evaluation of scientific performance as in the case of a football (soccer) league. The immediate observation that the famous scientists take the lead reinforces these suggestions.

A crucial question remains to be answered: how does the *h*-index relate to citation impact indicators based on advanced bibliometric indicators, and to the outcomes of peer review? This is particularly important because these advanced indicators do take into account the different citation patterns of the many disciplines and fields, but also of the different types of publications (e.g., 'normal' papers versus review articles). And certainly one of the *h*-index's 'main attractions' (Ball 2005) that it can rescue from obscurity those researchers who have made sustained and significant contributions but who have not won the reputation they deserve, is an attraction already provided for a long time by the advanced bibliometric indicators such as computed by our institute in Leiden. Therefore, it is time to compare the outcomes based on the *h*-index with advanced bibliometric indicators and of peer review. That there are advanced bibliometric indicators largely accepted by researchers and even preferred to peer assessment is apparently still not known well, given the remark in a very recent Nature editorial "Whether one is assessing individuals or their institutions, …everyone knows that most citation measures, while alluring, are overly simplistic. Unsurprisingly, most researchers prefer an explicit peer assessment of their work" (Nature 2005).

In this paper we address the above question at the level of research groups, i.e., statistically a level directly above that of the individual scientist. Thus, we deal with research groups rather than individual scientists, as we consider the research group as the most important work floor unit in research, particularly in the natural sciences. In most cases, however, the work of experienced, leading scientists closely approaches the oeuvre of their research group. In this sense we present unique material, as the research group is not an entity directly available in databases such as authors or journals. Research groups are defined by the internal structure of universities or other R&D institutions. We present characteristics of the statistical correlation between the *h*-index and several standard bibliometric indicators, as well as with the results of peer review judgment. We use the results of a large evaluation study covering all university research groups in chemistry and chemical engineering in the Netherlands. In particular, we focus on all publications of 147 chemistry research groups for the years 1991-1998, total number of publications about 18,000, and count the citations for a time window of three years starting with the publication year (i.e., for publications from 1991, citations are counted in the period 1991-1993, and for publications from 1998, citations are counted in the period 1998-2000). Only 'external' citations, i.e., citations corrected for self-citations, are taken into account.

2. Methodology and Data Material

The data set concerns all publications as far as published in journals covered by the Citation Index, 'CI publications'[2] of these 147 university research groups[3]. Thus, publications such as reports and books or book chapters are not taken into account. However, for chemistry research groups the focus on papers published in CI-covered journals generally provides a very good representation of the scientific output. In the framework of this evaluation study, we performed an extensive bibliometric analysis to support the international peer committee with quantitative evidences (van Leeuwen *et al* 2002). This enables us to calculate the *h*-index for all research groups and to compare this index with the standard bibliometric indicators already calculated for the evaluation study and, in addition, with the peer judgments. The peers used a three-point scale to judge the research quality of a group: Grade 5 is 'excellent', Grade 4 is 'good', and Grade 3 is 'satisfactory' (VSNU 2002). We present the standard bibliometric indicators with a short description in the text box here below, for details we refer to Van Raan (1996, 2004).

---

**Standard Bibliometric Indicators used for comparison with the *h*-index:**

- Number of publications (***P***) in CI-covered journals of the research group in the entire period;
- Number of citations received by ***P*** during the entire period, without self-citations (***C***);
- Average number of citations per publication, without self-citations (***CPP***);
- Journal-based worldwide average impact as an international reference level for the research group (***JCS***, journal citation score), without self-citations (on this world-wide scale!), in the case of more than one journal we use the average ***JCSm***; for the calculation of ***JCSm*** the same publication and citation counting procedure, time windows, and article type are used as in the case of ***CPP***;
- Field/subfield-based[4] worldwide average impact as an international reference level for the research group (***FCS***, (sub)field citation score), without self-citations (on this world-wide scale!) in the case of more than one (sub)field (as almost always) we use the average ***FCSm***; for the calculation of ***FCSm*** the same publication and citation counting procedure, time windows, and article type are used as in the case of ***CPP***;
- Comparison of the actually received international impact of the research group with the world-wide average based on ***JCSm*** as a standard, without self-citations, indicator ***CPP/JCSm***;
- Comparison of the actually received international impact of the research group with the world-wide average based on ***FCSm*** as a standard, without self-citations, indicator ***CPP/FCSm***;

---

[2] The former Institute for Scientific Information (ISI) in Philadelphia, now Thomson Scientific, is the producer and publisher of the Web of Science (WoS), which covers the Science Citation Index (extended version), the Social Science Citation Index, the Arts & Humanities Citation Index. Throughout this paper we use the term 'CI' (Citation Index) for the above set of databases.

[3] The (10) universities with chemistry departments covered by this evaluation study are Leiden, Utrecht, Groningen, Amsterdam UvA, Amsterdam VU, Nijmegen, Delft, Eindhoven, Enschede (Twente), and Wageningen.

[4] We here use the definition of (sub)fields based on a classification of scientific journals into *categories* developed by ISI/Thomson Scientific. Although this classification is not perfect, it provides a clear and 'fixed' consistent field definition suitable for automated procedures within our data-system.

A short explanation of the calculation of the indicators, particularly *CPP/FCSm*, is given in the appendix.

All fields within chemistry are covered by this set of university groups, the main fields being analytical chemistry, spectroscopy and microscopy; computational and theoretical chemistry, physical chemistry; catalysis; inorganic chemistry; organic and bio-organic chemistry; biochemistry, microbiology and biochemical engineering; polymer science and technology; materials science; chemical engineering. These fields may differ considerably in citation characteristics. Thus, for a fair comparison of impact of research groups in different fields it is essential to apply a field-specific normalization as it is indicated in the above indicator overview, particularly the indicator *CPP/FCSm*.

In Table 1 we show as an example the results of our analysis for the bibliometric indicators[5], the *h*-index[6] and the peer ratings for the twelve chemistry research groups of one of the ten universities ('Univ A'). In the calculation of the *h*-index we restrict the citation period to a three-year window instead of 'life time counts' in order to focus on the impact of recent work and thus on current research performance. The table makes clear that our indicator calculations allow an extensive statistical analysis of the correlation of both these indicators and peer judgment with the *h*-index for the entire set of research groups.

*Table 1: Example of the results of the bibliometric analysis for the chemistry groups*

| Research group | P | C | CPP | JCSm | FCSm | CPP/JCSm | CPP/FCSm | h-index | Quality |
|---|---|---|---|---|---|---|---|---|---|
| Univ A, 01 | 92 | 554 | 6.02 | 5.76 | 4.33 | 1.05 | 1.39 | 6 | 5 |
| Univ A, 02 | 69 | 536 | 7.77 | 5.12 | 2.98 | 1.52 | 2.61 | 8 | 4 |
| Univ A, 03 | 129 | 3780 | 29.30 | 17.20 | 11.86 | 1.70 | 2.47 | 17 | 5 |
| Univ A, 04 | 80 | 725 | 9.06 | 8.06 | 6.25 | 1.12 | 1.45 | 7 | 4 |
| Univ A, 05 | 188 | 1488 | 7.91 | 8.76 | 5.31 | 0.90 | 1.49 | 11 | 5 |
| Univ A, 06 | 52 | 424 | 8.15 | 6.27 | 3.56 | 1.30 | 2.29 | 9 | 4 |
| Univ A, 07 | 52 | 362 | 6.96 | 4.51 | 5.01 | 1.54 | 1.39 | 8 | 3 |
| Univ A, 08 | 171 | 1646 | 9.63 | 6.45 | 4.36 | 1.49 | 2.21 | 13 | 5 |
| Univ A, 09 | 132 | 2581 | 19.55 | 15.22 | 11.71 | 1.28 | 1.67 | 17 | 4 |
| Univ A, 10 | 119 | 2815 | 23.66 | 22.23 | 14.25 | 1.06 | 1.66 | 17 | 4 |
| Univ A, 11 | 141 | 1630 | 11.56 | 17.83 | 12.30 | 0.65 | 0.94 | 11 | 4 |
| Univ A, 12 | 102 | 1025 | 10.05 | 10.48 | 7.18 | 0.96 | 1.40 | 10 | 5 |

---

[5] The standard bibliometric indicators are calculated with a 'total block analysis', which means that publications are counted for the entire 10-year period from 1991-2000 and citations are counted up to and including 2000 (e.g., for publications from 1991, citations are counted in the period 1991-2000, and for publications from 2000, citations are counted only in 2000).
[6] In order to focus on recent performance, the *h*-index is calculated with *fixed citation window*: we take the publications of the chemistry research groups for the years 1991-1998 and count the citations for a time window of three years starting with the publication year (i.e., for publications from 1991, citations are counted in the period 1991-1993, and for publications from 1998, citations are counted in the period 1998-2000).

According to Hirsch, several of the above standard indicators are 'commonly used' (particularly *P*, *C*, *CPP*) and they have a number of disadvantages. The *h*-index is supposed to measure the broad impact of an individual scientist and to avoid all the disadvantages. Moreover, it can usually be found very easily -for individual scientists- in the Web of Science (WoS). We comment as follows. First, the total number of publications (*P*) measures productivity, and not impact. It is not difficult to agree with that. Second, the total number of citations (*C*) and the number of citations per paper (*CPP*) would be hard to find. This may be true for researchers dependent on the WoS, but for bibliometric research groups these indicators must be -and they are- directly available from their data system. Moreover, bibliometric research groups put much effort in cleaning data in order to correct for many possible errors in names of individual scientists, and at the level of research groups these problems multiply. These technical problems are discussed in Van Raan (2005) and Moed (2005).

Methodologically more important are the arguments of Hirsch that *C* may be inflated by a small number of 'big hits' which may not be representative of the individual scientists if he/she is coauthor with many others on those 'big hit' papers, that *C* also gives undue weight to highly cited review articles versus original research papers, and that *CPP* -though allowing comparison of scientists of different ages- rewards low productivity and penalizes high productivity. In our opinion, the 'big hit' problem certainly may exist a soon as coauthors are involved. However, at the level of research groups this problem will be less problematic. Review papers do indeed attract, on the average, considerably more citations than 'normal' papers, but as discussed in our indicator overview, we solved this problem by using 'article-type normalized' indicators. With regard to low or high productivity, we stress that scientists as well as research groups differ considerably in terms of number of publications because of differences in field and type of research (for instance, theoretical versus experimental groups).

The number of papers with more than *y* citations, or the number of citations to each of the *q* (e.g., 5) most cited papers would eliminate, according to Hirsch, the supposed disadvantages connected to *P*, *C*, and *CPP*. The disadvantage is the arbitrariness of the values of *y* or *q*, which will randomly favor or disfavor individual scientists and they need to be adjusted for seniority. Furthermore, *q* is not a single number, which makes comparisons more difficult.

Because of these various aspects of research groups, we emphasize the importance of the use of a set of different indicators and field-normalization. In the next section we focus our comparison of the *h*-index and standard bibliometric indicators on our field- and article-type normalized 'crown indicator' ***CPP/FCSm***.

## 3. Results and Discussion

We first present the main empirical findings of this study: the correlation of the *h*-index with the total number of citations (Fig. 1) and with the total number of publications (Fig. 2); the correlation of our crown indicator *CPP/FCSm* with the total number of citations (Fig. 3) and with the total number of publications (Fig. 4); the correlation of the *h*-index with ***CPP/FCSm*** (Fig. 5); and finally the correlation of the *h*-index and ***CPP/FCSm*** with peer judgment in Table 2.

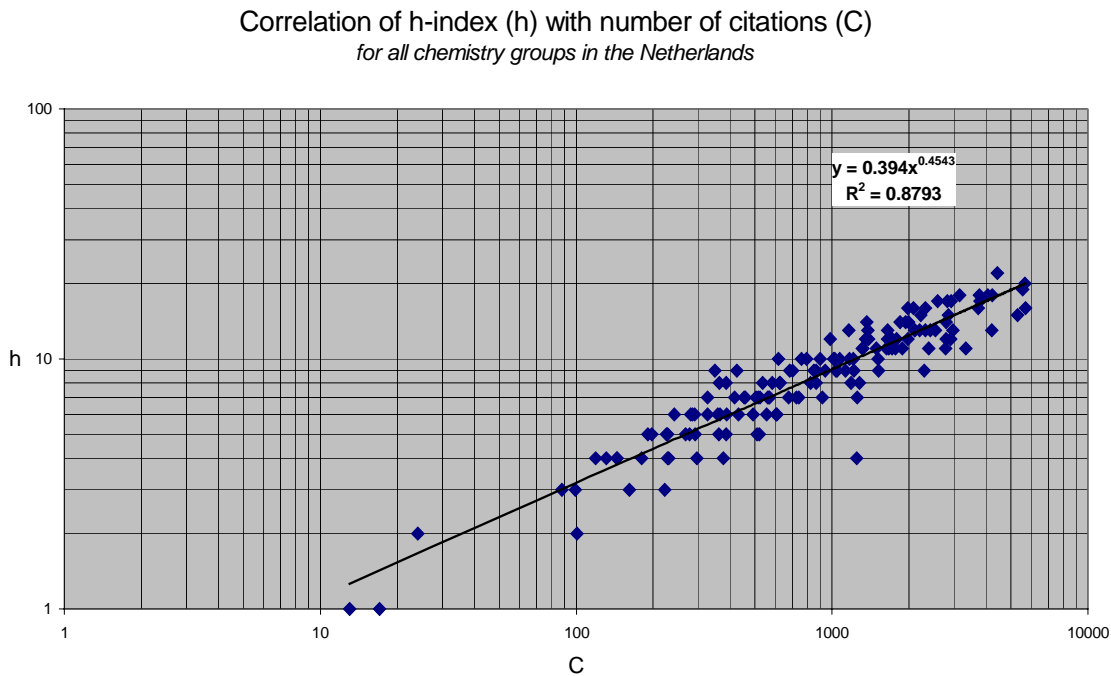

*Figure 1*: *Correlation of the h-index with the total number of citations for all chemistry research groups*

We observe in Fig. 1 a very good correlation ($R^2 = 0.89$) between the *h*-index and the total number of citations (***C***) for all 147 chemistry research groups and find the following relation:

***h*** = 0.42 ***C***$^{0.45}$

which means that $h^2$ is approximately an order of magnitude lower than the total number of citations of a research group, as can be easily seen in Fig. 1. Because our calculation of the *h*-index is based on a three-year window, this relation between the values of $h^2$ and ***C*** is to be expected.

Figure 2 shows the correlation between the *h*-index and the total number of publications for all 147 chemistry research groups. This correlation is less strong than in the case of citations.

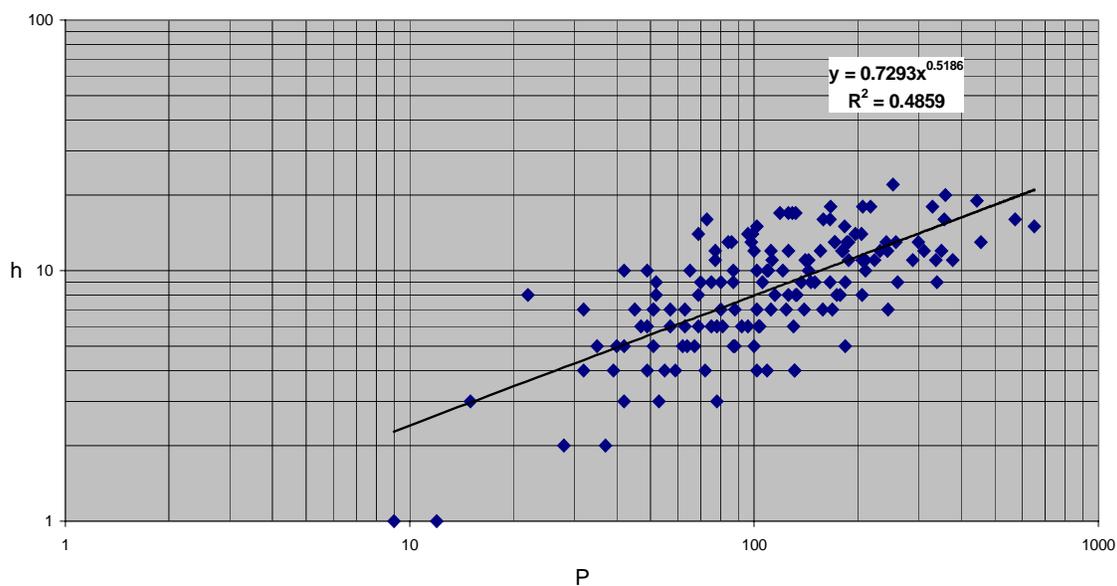

*Figure 2*: *Correlation of the h-index with the total number of publications for all chemistry research groups*

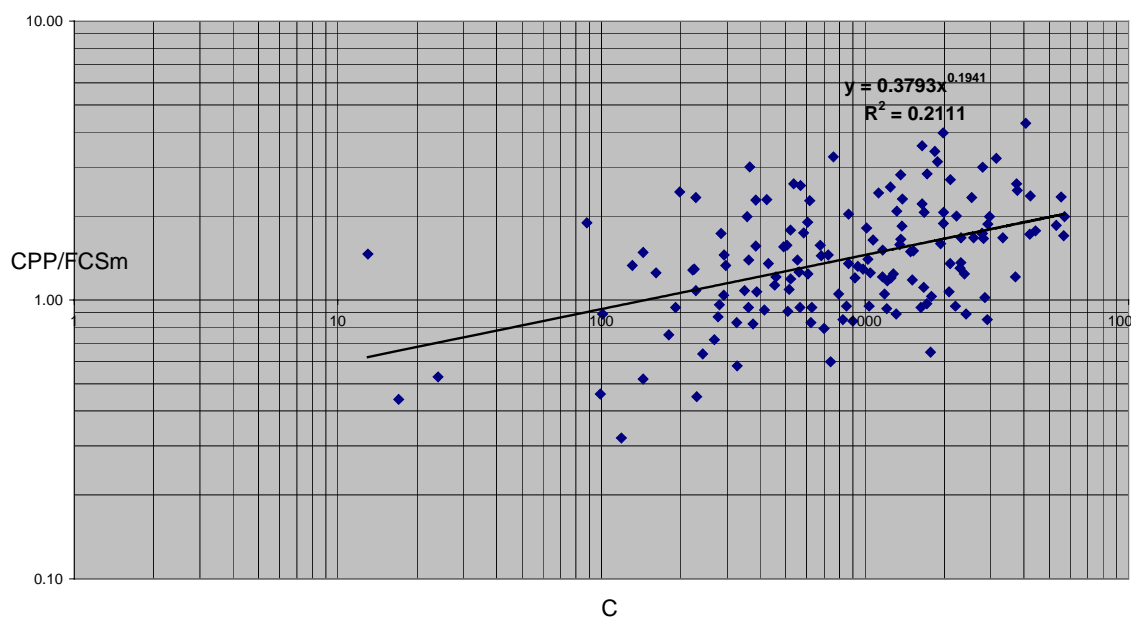

*Figure 3*: *Correlation of the crown indicator (CPP/FCSm) values with the total number of citations for all chemistry research groups*

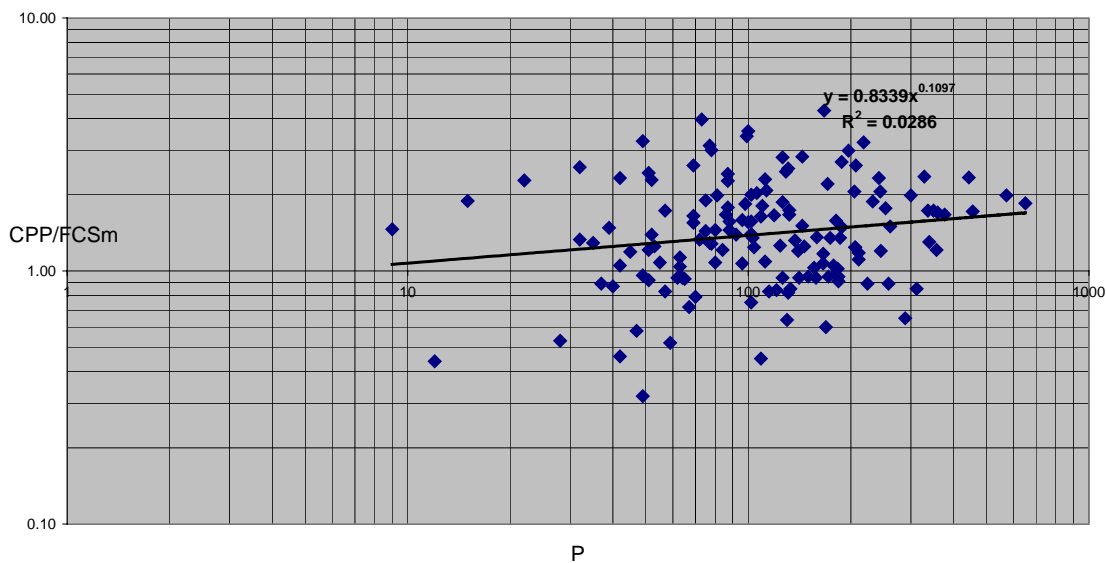

*Figure 4*: *Correlation of the crown indicator (CPP/FCSm) values with the total number of publications for all chemistry research groups*

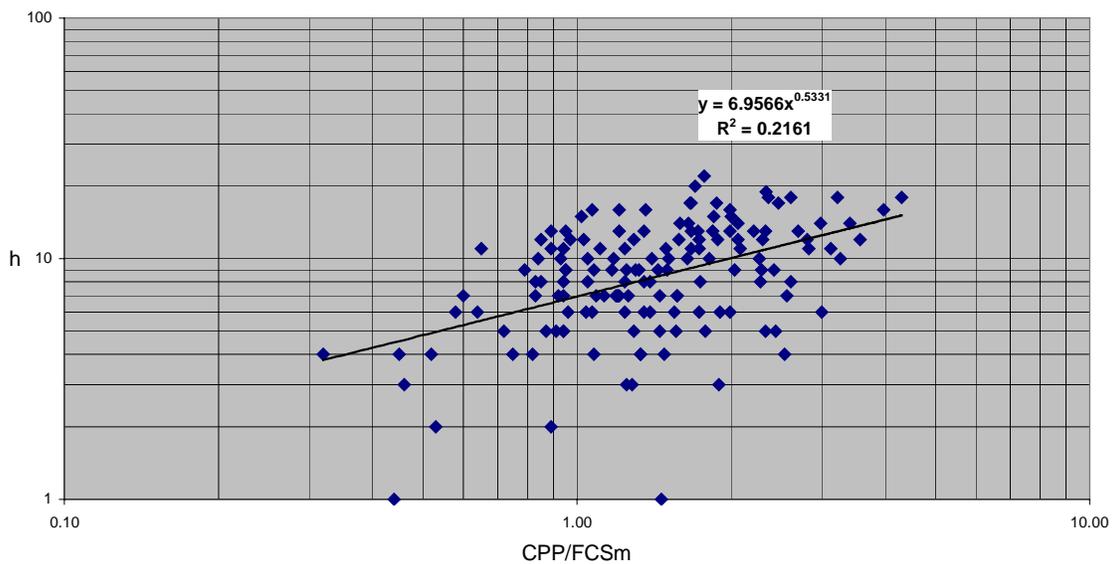

*Figure 5*: *Correlation of the h-index with the crown indicator (CPP/FCSm) values for all chemistry research groups*

In contrast to the findings for the *h*-index, we observe that the significance of the correlation of our crown indicator ***CPP/FCSm*** with the total number of citations (Fig. 3) is very small, and with the number of publications (Fig. 4) practically non-existent. In Fig. 5 we present the correlation between the *h*-index and ***CPP/FCSm***. We observe that this correlation is very low. These results mean that an important difference between the *h*-index and our crown indicator is *size-dependence*, as can be expected given the definition of these indicators. For the *h*-index 'big' is important, but for our crown indicator small can also be beautiful.

Finally, in Table 2 we present the relation between the *h*-index as well as the ***CPP/FCSm*** indicator with peer judgment for all chemistry groups. We use the results of Fig. 5 to determine classes of *h*-index values that are reasonably comparable with classes of ***CPP/FCSm*** values.

| Q | 3 | 4 | 5 | sum |
|---|---|---|---|---|
| *h* | | | | |
| >10 | 1 | 30 | 26 | *57* |
| 7 to 10 | 11 | 32 | 6 | *49* |
| < 7 | 18 | 16 | 7 | *41* |
| | | | | |
| sum | *30* | *78* | *39* | **147** |

| Q | 3 | 4 | 5 | sum |
|---|---|---|---|---|
| *CPP/FCSm* | | | | |
| > 2.00 | 1 | 16 | 15 | *32* |
| 1.00 to 2.00 | 15 | 44 | 20 | *79* |
| ≤ 1.00 | 14 | 18 | 4 | *36* |
| | | | | |
| sum | *30* | *78* | *39* | **147** |

*Table 2*: *Relation between h-index classes (left hand side) and CPP/FCSm classes (right hand side) and peer judgment (Q)*

We clearly observe that research groups with a high *h*-index (*h* > 10) are about evenly distributed between peer judgment ratings 4 and 5, and the same is the case for the highest ***CPP/FCSm*** class. Thus, both indicators discriminate very well between research groups rated excellent (***Q*** = 5) or good (***Q*** = 4) on the one side, and less good ('satisfactory', ***Q*** =3) on the other, but less well between good and excellent. This finding is in line with observations of Moed (2005) concerning the relation between bibliometric indicators and peer ratings.

The first conclusion is that the *h*-index and the crown indicator both relate in a quite comparable way with peer judgments. Given the distribution between *h*-index classes and ***CPP/FCSm*** classes for ***Q*** = 4, ***CPP/FCSm*** relates better to the peer judgment than the *h*-index. By analyzing the data for more extreme cases, we find among the research groups within the *top*-25% of the *h*-index distribution 1 group with a less good peer rating, which means a *high h*-value but a *low* peer rating (***Q*** = 3). This group has ***CPP/FCSm*** = 1.02, whereas the average ***CPP/FCSm*** value for all around 40 groups in the top-25% of the *h*-index is 2.10. Thus, here our crown indicator better reflects the lower performance of this group than the *h*-index.

For the research groups belonging to the *bottom*-25% of the *h*-index distribution we find as many as 7 groups with an excellent peer rating, which means a *low h*-value but a *very* *high* peer rating ($Q$ = 5). These (small) groups have a **CPP/FCSm** = 1.70, whereas the average **CPP/FCSm** value for all groups in the bottom-25% of the *h*-index is 1.25. Now our crown indicator better reflects the high performance of these groups than the *h*-index.

## 4. Concluding Remarks

In most cases, peers are very well aware of highly productive groups with considerable scientific impact in their field. Because the *h*-index indicates 'brute force in citations' we indeed can expect a significant correlation with peer judgment. We will find the same by simply using the total number of citations (*C*), given the very strong correlation between the *h*-index and *C*, as illustrated by Fig. 1. But the situation is different for smaller groups in fields with 'less heavy citation traffic'. As shown above, particularly for this type of research groups our crown indicator is, in our opinion, a more appropriate indicator of research performance.

It has also been stressed as a fortunate aspect that the *h*-index 'happily ignores' the journal impact factor (Nature 2005). But should we really be so happy? First, there are journal-based indicators that are more suited for evaluation purposes than the common impact factor. The **JCSm** indicator discussed in this paper is a long-standing example. Second, precisely the aspect of disregarding the journal may severely limit the usefulness of the *h*-index because of the possibilities of manipulation. For instance, if in a specific journal of mediocre or rather low level, groups of researchers deliberately start citing overly each other's work, they 'artificially' increase their *h*-indexes. With our advanced bibliometric indicators this manipulation would be observed immediately by comparison of the journal-normalized indicator with the other indicators. Again, it is not wise to force the assessment of researchers or of research groups into just one specific measure. It is even dangerous, because it reinforces the opinion of administrators and politicians that scientific performance can be expressed simply by one note. That is why we always stress that a consistent set of several indicators is necessary, in order to illuminate different aspects of performance.

*Acknowledgements*

The author thanks his CWTS colleague Dr Thed van Leeuwen for the bibliometric data analysis in the chemistry research groups' performance study.

*Appendix: Calculation of CPP/JCSm and CPP/FCSm*

We take as an example the 'oeuvre' of a research group consisting of four publications (I, II, III, IV) of different article type in the given journals belonging to specific fields, time period 1996-1999. In the table we give the indicator **CPP** is given for each publication (**P** =1, thus in these case **CPP** is simply the number of citations up to 1999). For each of the four publications separately, a **JCS** and **FCS** value is calculated. These values are the **CPP**-values for the entire journal and field (sets of journals), respectively, belonging to that specific publication of the group, taking into account the same citation period (so in case of publication I it is the period 1996-1999, for II it is 1997-1999, and for III and IV it is only the year 1999), and the same article type in the journal and field concerned.

| article type | publ. year | journal | field | cit. up to 1999 = CPP | JCS | FCS |
|---|---|---|---|---|---|---|
| I *review* | 1996 | *CANCER RES* | Oncology | 17 | *16.9* | *23.7* |
| II *note* | 1997 | *J CLIN END* | Endocrinology | 4 | *3.1* | *3.0* |
| III *article* | 1999 | *J CLIN END* | Endocrinology | 6 | *4.8* | *4.1* |
| IV *article* | 1999 | *J CLIN END* | Endocrinology | 8 | *4.8* | *4.1* |

Calculation of the bibliometric indicators **CPP**, **JCSm**, and **FCSm** for the entire 'oeuvre' of four publications for the period 1996-1999:

**CPP**  = (17+4+6+8) / (1+1+1+1) = 8.75
**JCSm** = [(1 x 16.9)+(1 x 3.1)+(2 x 4.8)] / (1+1+2) = 7.40
**FCSm** = [(1 x 23.7)+(1 x 3.0)+(2 x 4.1)] / (1+1+2) = 8.72

Hence, **CPP/JCSm** = 8.75 / 7.40 = 1.18, and **CPP/FCSm** = 8.75 / 8.72 = 1.00.

version 28 02 06